\documentclass[12pt]{iopart}

\usepackage{amsmath}
\usepackage{amssymb}
\usepackage{hyperref}

\newcommand{\Msun}{M_\odot}
\newcommand{\Asharp}{A^\sharp}

\usepackage[backend=biber,style=ieee,sorting=none]{biblatex}
\usepackage{extsizes}
\usepackage[margin=1in]{geometry}
\usepackage{graphicx}
\usepackage{caption}
\hypersetup{
    colorlinks=true,
    linkcolor=blue,
    citecolor=blue,
    filecolor=blue,      
    urlcolor=blue,
}

\begin{document}

\title[Measuring Neutron Star Radius with 2G \& 3G GW Detectors]{Measuring Neutron Star Radius with second and third generation Gravitational Wave Detector Networks}

\author[0000-0002-5116-844X]{Ananya Bandopadhyay}
\ead{abandopa@syr.edu}
\address{Department of Physics,
Syracuse University, 
Syracuse, NY 13244, USA}

\author[0009-0004-9167-7769]{Keisi Kacanja}
\ead{kkacanja@syr.edu}
\address{Department of Physics,
Syracuse University, 
Syracuse, NY 13244, USA}

\author[0000-0003-0427-3893]{Rahul Somasundaram}
\ead{rsomasun@syr.edu}
\address{Department of Physics,
Syracuse University, 
Syracuse, NY 13244, USA}
\address{Theoretical Division, Los Alamos National Laboratory, Los Alamos, NM 87545, USA}

\author[0000-0002-1850-4587]{Alexander H. Nitz}
\ead{ahnitz@syr.edu}
\address{Department of Physics,
Syracuse University, 
Syracuse, NY 13244, USA}

\author[0000-0002-9180-5765]{Duncan A.~Brown}
\ead{dabrown@syr.edu}
\address{Department of Physics,
Syracuse University, 
Syracuse, NY 13244, USA}

\vspace{10pt}
\begin{indented}
\item[]February 2023
\end{indented}

\begin{abstract}
The next generation of ground-based interferometric gravitational wave detectors will observe mergers of black holes and neutron stars throughout cosmic time. A large number of the binary neutron star merger events will be observed with extreme high fidelity, and will provide stringent constraints on the equation of state of nuclear matter. In this paper, we investigate the systematic improvement in the measurability of the equation of state with increase in detector sensitivity by combining constraints obtained on the radius of a $1.4 \, \mathrm{M}_{\odot}$ neutron star from a simulated source population. Since the measurability of the equation of state depends on its stiffness, we consider a range of realistic equations of state that span the current observational constraints. We show that a single 40km Cosmic Explorer detector can pin down the neutron star radius for a soft, medium and stiff equation of state to an accuracy of 10m within a decade, whereas the current generation of ground-based detectors like the Advanced LIGO-Virgo network would take $\mathcal{O}(10^5)$ years to do so for a soft equation of state. 

\end{abstract}

\section{Introduction} \label{sec:intro}

Cosmic Explorer is a groundbreaking next-generation gravitational wave detector, representing a significant leap in our capacity to explore a plethora of compact binary merger events. The development of Cosmic Explorer and other advanced gravitational wave detectors, such as Einstein Telescope~\cite{Hild:2009ns,Sathyaprakash:2012jk}, will allow us to understand neutron stars and their internal structure. Cosmic Explorer will detect hundreds of signals with high signal-to-noise ratios (SNR), greater than 100~\cite{LIGOScientific:2016wof,2021arXiv210909882E}. With the improvement in sensitivity of the detectors, it will be possible to measure the radius of a neutron star with unprecedented accuracy. Inference from simulated populations performed by combining information from a large number of observed events, allows us to obtain improved constraints on the equation of state~\cite{Lackey:2014fwa,Agathos:2015uaa,HernandezVivanco:2019vvk,Pacilio:2021jmq,Chatziioannou:2021tdi,Finstad23}. Previous works have demonstrated the feasibility of combining constraints from multiple observations using statistical tools, to obtain a tighter constraint on the equation of state.

In particular, Finstad et al.~\cite{Finstad23} combined $\mathcal{O}(300)$ signals in an Advanced LIGO-Virgo network and compared it to the constraints obtained by stacking observations for one year of operation of a single Cosmic Explorer detector, showing the order of magnitude improvement in measurement accuracy achieved by a singular Cosmic Explorer. 
However, Cosmic Explorer will operate as a network which calls for the need to systematically investigate different detector configurations in order to fully gauge the improvement in sensitivity that Cosmic Explorer would accomplish. Therefore, in this paper, we extend the work of Finstad et al. ~\cite{Finstad23} to systematically study the improvement in the measurability of the equation of state with the increase in network sensitivity from the current generation of Advanced LIGO-Virgo detectors at design sensitivity to the planned Cosmic Explorer detectors, which are expected to begin operations in 2030s. We simulate an astrophysical population of neutron stars consistent with current observational constraints on the rate of binary neutron star mergers in the universe~\cite{LIGOScientific:2020kqk}. We perform a full Bayesian parameter estimation to obtain a forecast on the expected number of years of observation at different detector sensitivities required to measure the radius of a neutron star accurate to 10m, by simulating different realizations of the universe from the above population. We present our results for an Advanced LIGO-Virgo network, a post O5 network comprised of three LIGO detectors at $\rm \Asharp$ \footnote{\href{https://dcc.ligo.org/LIGO-T2300041/public}{https://dcc.ligo.org/LIGO-T2300041/public}} sensitivity, a 20km and a 40km Cosmic Explorer detector \footnote{\href{https://dcc.cosmicexplorer.org/CE-T2000017}{https://dcc.cosmicexplorer.org/CE-T2000017}}, and two other next generation detector networks. The first of these is comprised of a 40km Cosmic Explorer along with two detectors at $\rm \Asharp$ sensitivity, and the second consists of a 40km Cosmic Explorer, a 20km Cosmic Explorer and one detector at $\rm \Asharp$ sensitivity. 
For each of the networks, we combine equation of state constraints (in terms of measurement of radius of a $1.4 \, \rm M_{\odot}$ neutron star) from multiple events, observed across a given observing period for each detector (network). We study the measurability of the neutron star radius for three different equation of state models drawn from ~\cite{Capano:2019eae} that are calibrated with nuclear theory at lower densities and span the current observational constraints at densities relevant for neutron stars. We find that, while the current generation of ground-based gravitational wave detector network would require thousands of years to accurately measure the neutron star radius even for a stiff equation of state (i.e. relatively more measurable), a single 40km Cosmic Explorer can do so within a year. Multi-detector networks including at least one Cosmic Explorer detector could constrain the radius to 10m accuracy within a few years for even the softest equation of state considered here. Our result clearly demonstrates the potential of next generation detector networks to constrain the equation of state of dense nuclear matter to a higher accuracy than is possible for any existing experimental facility.  

The rest of this article is organized as follows. In Section~\ref{sec:population}, we describe the properties of the population of neutron stars that we are examining for our study. In Section~\ref{sec:methods}, we describe the Bayesian inference framework that is used to obtain combined constraints on the equation of state from the measurement of neutron star radius across an observed population. In Section~\ref{sec:results}, we present the results of our analysis, i.e. the projected constraints on the radius of a $1.4 \, \rm M_{\odot}$ neutron star for different detector networks, and conclude the study in Section~\ref{sec:conclusion}.

\section{Neutron Star Population} \label{sec:population}

To study the measurability of tidal deformability from the gravitational wave signals of inspiralling binary neutron star systems, we generate a population of binary neutron star systems with component masses drawn from a Gaussian centered at $1.4 \, \rm M_{\odot}$ and having a standard deviation $\sigma_m = 0.05 \, \rm \Msun $. The spins are chosen to be aligned with the orbital angular momentum, and are drawn from a Gaussian distribution with zero mean and $\sigma_{\chi} = 0.02.$ The events are chosen to be isotropically distributed across sky locations, and the inclination and orientation of the binary systems are distributed uniformly on the sphere. 
The luminosity distances of the systems are drawn from a uniform distribution in the range $d_L\in [20 \, \rm Mpc,20 \, \rm Gpc].$ Events at larger distances, i.e. having low values of signal-to-noise ratios, contain roughly the same amount of information about the measured parameters, thus drawing events uniformly in distance reduces the amount of computational resources spent on sampling over the parameters of a large number of ``uninformative'' events occurring at higher distances in the simulated astrophysical population. It also gives us a greater selection of ``informative'' events (occurring at nearby distances) to choose from, in constructing different realizations of the universe, as compared to what we would have had if our simulated population of neutron stars was the true astrophysical population. We use the median merger rate for binary neutron stars $\mathcal{R}=320_{-240}^{+490}\,\rm Gpc^{-3} yr^{-1}$  from~\cite{LIGOScientific:2020kqk} to normalize the astrophysical population, and reweight our neutron star population to a uniform in volume distribution. Our final distribution is uniform in volume. The uniform in volume assumption is reasonable for sources which contribute substantially to the equation of state measurement, since only the close by sources would be detected with signal-to-noise ratios high enough to extract equation of state information from them.

\begin{figure}[t]
    \begin{center}
    \advance\leftskip-0.5cm
    \includegraphics[height=9cm,width=13cm]{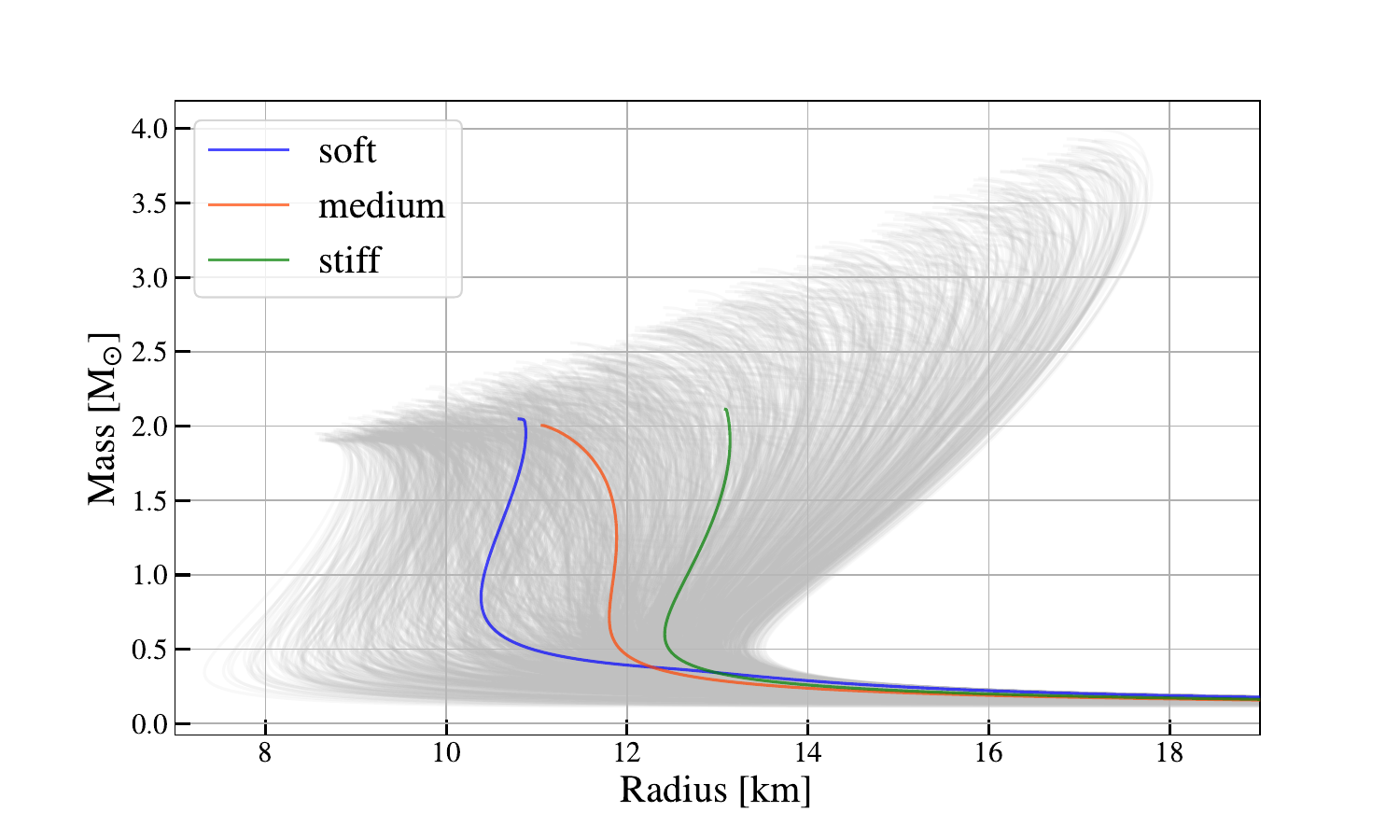}
    \caption{Neutron star mass-radius curves for the soft (blue), medium (orange) and stiff (green) equations of state used in our analysis. The curves shown in grey depict the full set of 2000 equations of state used as the prior for the Bayesian analysis. The equation of state set is constructed in such a way that the distibution of radii for a $1.4 \, \rm M_{\odot}$ neutron star is roughly uniform. }
    \label{mass-radius-curves}
    \end{center}
\end{figure} 

Since the measurability of the tidal deformability depends strongly on the equation of state, we choose a set of three equations of state representing a soft, a medium and a stiff one, for this study. These are then used to assign tidal deformabilities to the individual neutron stars in each simulated signal. They are chosen from a set of 2000 equations of state also used in ~\cite{Finstad23}, calibrated with chiral effective field theory calculations up to nuclear saturation density, $n_{\rm sat}$. Beyond this point, their construction is based on general considerations of thermodynamic stability and causality, and employing a general extension scheme in the speed of sound  $c_{\rm s}^2$ for densities between $[n_{\rm sat},12n_{\rm sat}]$~\cite{Capano:2019eae}. Fig.~\ref{mass-radius-curves} shows the full set of 2000 equations of state, which we use as the prior for our parameter estimation runs, with the soft, medium and stiff equations of state labelled in blue, green and orange respectively.\\
Parameter estimation runs are performed for the simulated population separately for each equation of state. The waveform model used for generating the simulated signals is the \texttt{IMRPhenomD\_NRTidal} approximant. This is a frequency domain approximant that is based on a phenomenological inspiral-merger-ringdown model \texttt{IMRPhenomD},  originally constructed to model binary black hole systems with spins aligned with their angular momentum axis~\cite{Husa:2015iqa,Khan:2015jqa}. Tidal effects are superposed on the binary black hole waveforms and calibrated with Numerical Relativity simulations of binary neutron star systems to obtain the binary neutron star waveform approximant~\cite{Barkett:2015wia,Dietrich:2017aum}. For high SNR signals, which will be detected abundantly by the next generation gravitational wave detectors, waveform systematics can become a dominant source of error in the extraction of parameters~\cite{Chatziioannou:2021tdi}. Earlier works have shown that tidal parameters can be recovered reliably using the \texttt{IMRPhenomD\_NRTidal} waveform for systems with small mass ratios and spins~\cite{Dietrich:2020eud}. Stiffer equations of state have been shown to introduce a bias in the recovery of non-tidal parameters, which in turn would affect the measurement of the equation of state~\cite{Dudi:2018jzn,Dietrich:2020eud}. Systems with unequal mass ratios and highly spinning systems also show an increased bias in the estimation of tidal parameters~\cite{Samajdar:2018dcx}, thus indicating the need for improved waveform modelling for binary neutron star systems for the next generation of ground based detectors.

\section{Injection Study} \label{sec:methods}
To quantify the improvement in measurability of the equation of state through neutron star radius and tidal deformability measurements with increase in detector sensitivity, we perform a Bayesian Inference based parameter estimation for the injected signals, which are projected onto simulated detector noise. We consider six different detector networks for the purpose of this study. The LIGO-Virgo network consists of the LIGO Handford, LIGO Livingston~\cite{LIGOScientific:2016emj,2020PhRvD.102f2003B} and Virgo detectors~\cite{VIRGO:2014yos} simulated at design sensitivity~\cite{KAGRA:2013rdx}. The post-O5 network is also comprised of three detectors, LIGO Handford, LIGO Livingston and LIGO India, at A$^\sharp$ sensitivity~\cite{T2200287}. The two Cosmic Explorer detectors are considered as single detector networks, a 40km Cosmic Explorer detector off the coast of Washington, and a 20km Cosmic Explorer detector off the coast of Texas. Additionally, we consider two more next generation detector networks. The first of these is comprised of a 40km Cosmic Explorer detector off the coast of Washington, and two detectors at Livingston and India, operating at A$^\sharp$ sensitivity (henceforth referred to as CE40+2$\rm \Asharp$). The final network consists of a 40km Cosmic Explorer, a 20km Cosmic Explorer and LIGO India operating at A$^\sharp$ sensitivity (henceforth referred to as CE40+CE20+$\rm \Asharp$).  Since the site locations for the Cosmic Explorer detectors are yet to be determined, the aforementioned fiducial locations are chosen as being close enough to a wide range of potential sites for the detectors to be representative from the point of view of the science goals that we address here. \\
To assess the performance of each detector network, we simulate a population of binary neutron star sources and apply a lower bound of 30 on the signal-to-noise ratio (SNR) to determine the events for which parameter estimation runs are performed to infer source parameters. For signals observed with lower signal-to-noise ratio than the threshold value, the error bars associated with radius measurement are large, implying that these signals contribute negligibly to the combined equation of state inference.\\
Parameter estimation for the injected signals is based on the Bayes theorem. Here we briefly review the theorem in the context of its application to gravitational wave astronomy. It states that for a set of parameters $\vec{\theta}$ that can be used to construct a model $M$ for the gravitational waveform, and given $\vec{d}(t)$, the gravitational wave strain data from the detectors,
\begin{equation}
    p(\vec{\theta}|\vec{d}(t),M) = \frac{\pi(\vec{\theta}(t)|M) \mathcal{L}(\vec{d}(t)|\vec{\theta},M)}{\mathcal{Z}},
\end{equation}
where $p(\vec{\theta}|\vec{d}(t),M)$ is the data-informed posterior distribution of the parameter $\vec{\theta}$, $\pi(\vec{\theta}(t)|M)$ is our prior knowledge of the parameter vector $\vec{\theta}$, conditional on the model, and $\mathcal{L}(\vec{d}(t)|\vec{\theta},M)$ is the likelihood of obtaining data $\vec{d}(t)$ given the waveform model $M$ and the parameters $\vec{\theta}$. Assuming that each detector produces stationary, Gaussian noise that is uncorrelated between different detectors in the network, the likelihood function is given by~\cite{1970esn..book.....W}
\begin{align}
    \mathcal{L}(\vec{d}|\vec{\theta},M) =\exp \left( -\frac{1}{2} \sum \limits_{i=1}^N\langle n_i|n_i\rangle\right) \nonumber \\
    =\exp \left( -\frac{1}{2} \sum \limits_{i=1}^N \langle d_i-h_i(\vec{\theta})|d_i-h_i(\vec{\theta})\rangle\right)
\end{align}

where
\begin{equation}
    \langle a|b \rangle = 4 \mathbb{R}\int\limits_{f_{\rm min}}^{f_{\rm max}}{\frac{\tilde{a}^*(f) \tilde{b}(f)}{S_n(f)}}
\end{equation}

is the noise-weighted inner product of two functions.

$\mathcal{Z} = \int {\pi(\vec{\theta}(t)|M) \mathcal{L}(\vec{d}(t)|\vec{\theta},M)d\vec{\theta}} $ is the Bayesian evidence, which acts as a normalization constant. We use ensemble samplers, like Markov Chain Monte Carlo samplers, that sample the parameter space of interest, and trace out the posterior probability distribution. For this work, we use the parallel-tempered \texttt{emcee} sampler~\cite{Foreman-Mackey:2012any}, invoking it through the \texttt{PyCBC Inference} library~\cite{Biwer:2018osg}, to estimate posterior distributions of all of the parameters used to model the binary neutron star signals, as described in Section~\ref{sec:population}. We use the heterodyne likelihood model~\cite{Cornish:2021lje,Zackay:2018qdy,Finstad:2020sok} in \texttt{PyCBC} for likelihood estimation, which evaluates the likelihood in discrete bins close to the peak value, to speed-up sampler convergence. \\
In the process of recovering the signal parameters through Markov-Chain Monte Carlo sampling, the sampling parameters are chosen to be the source frame chirp mass and mass ratio, component spins along the direction of the angular momenta, sky location, distance, geocentric time of coalescence, inclination, polarization angle, and the equation of state. 
We sample over the equations of state in the same procedure as ~\cite{Finstad23}. The equations of state are indexed in increasing order of radii for a $1.4 \, \rm M_{\odot}$ neutron star, and each index corresponds to a unique mass-radius (or equivalently, mass-tidal deformability) curve in the equation of state parameter space.
The lower frequency cutoff for the signals analyzed are chosen to be higher than the seismic frequency limit for the detectors. For the LIGO-Virgo network at design sensitivity, we use a lower frequency cutoff of 20~Hz, for the detectors at $\rm \Asharp$ sensitivity, we use 10~Hz, and for the Cosmic Explorer detectors, the lower frequency cutoff is chosen to be 7~Hz. The high frequency cutoff for all signals is 2048~Hz. \\
Once the sampler converges, we obtain the full set of posterior samples for the parameters of each of the analyzed events. This represents our knowledge of the source parameters for the gravitational wave event, conditioned on the data. We then combine multiple events hierarchically, to obtain a hyperposterior for the neutron star radius, that represents our integrated knowledge of the equation of state from the observed population of neutron stars. To do this, we first convert the equation of state samples to radius samples for a $1.4 \, \rm M_{\odot}$ neutron star (denoted as $\rm R_{1.4}$). This is done by using the mass-radius curve corresponding to each equation of state sample to determine the radius value for that sample. We then use a Gaussian Kernel Density Estimation~\cite{10.1214/aoms/1177728190} to obtain the posterior distribution on the radius for each event. The combined posterior for $N$ events $\lbrace s_1,s_2,...s_N\rbrace$ is then obtained as
\begin{equation}
    p(\mathrm{R}_{1.4}|s_1,...s_N) = p(\mathrm{R}_{1.4})^{(1-N)}\prod \limits_{i=1}^{N}p(\mathrm{R}_{1.4}|s_i) \label{combinedposteriors}
\end{equation}
where $p(\mathrm{R}_{1.4})$ is the radius prior, uniform across all events. 
\begin{figure*}
    \advance\leftskip-1cm
    \includegraphics[height=6cm,width=0.495\linewidth]{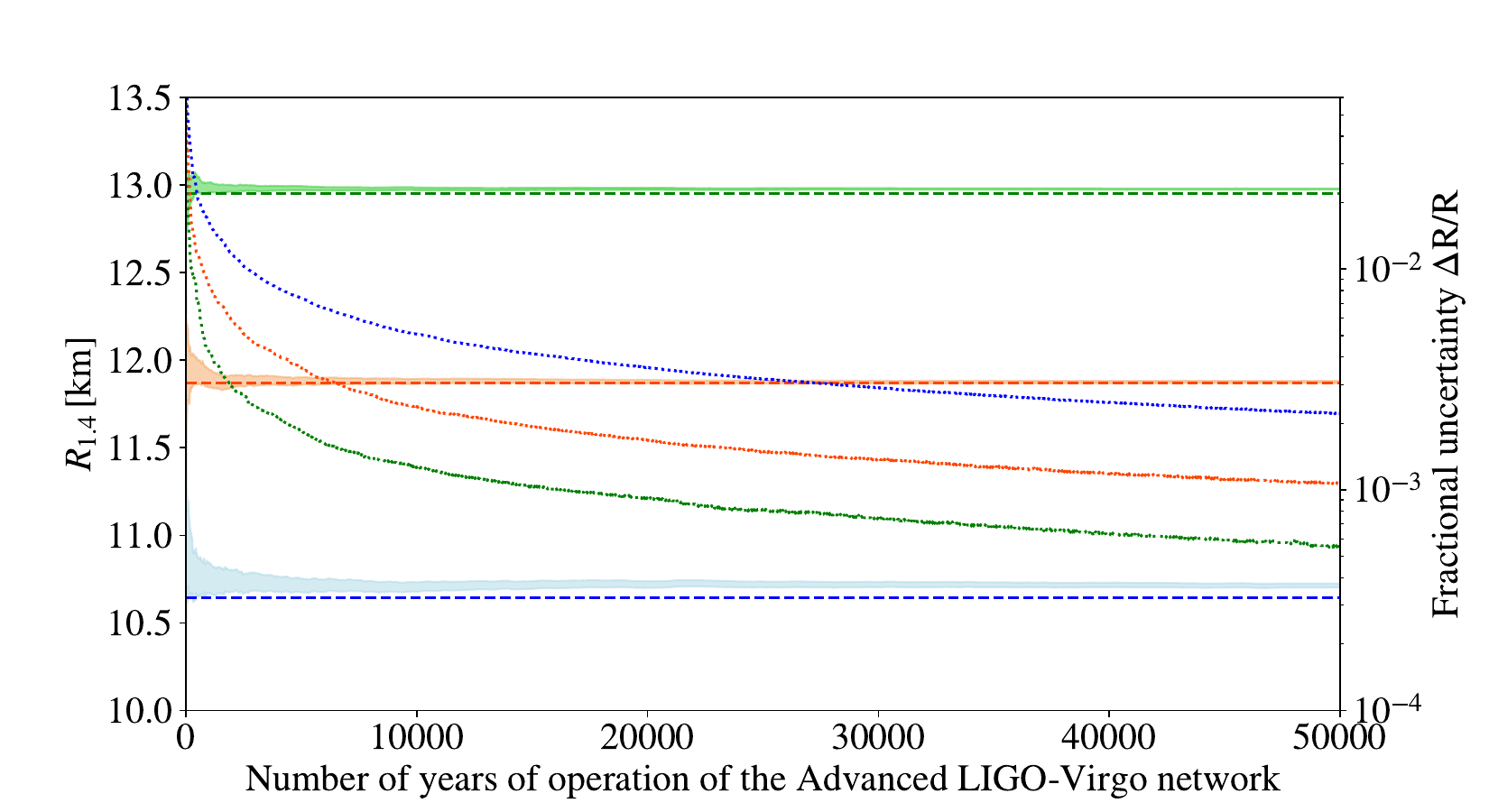}\hfill
    \includegraphics[height=6cm,width=0.495\linewidth]{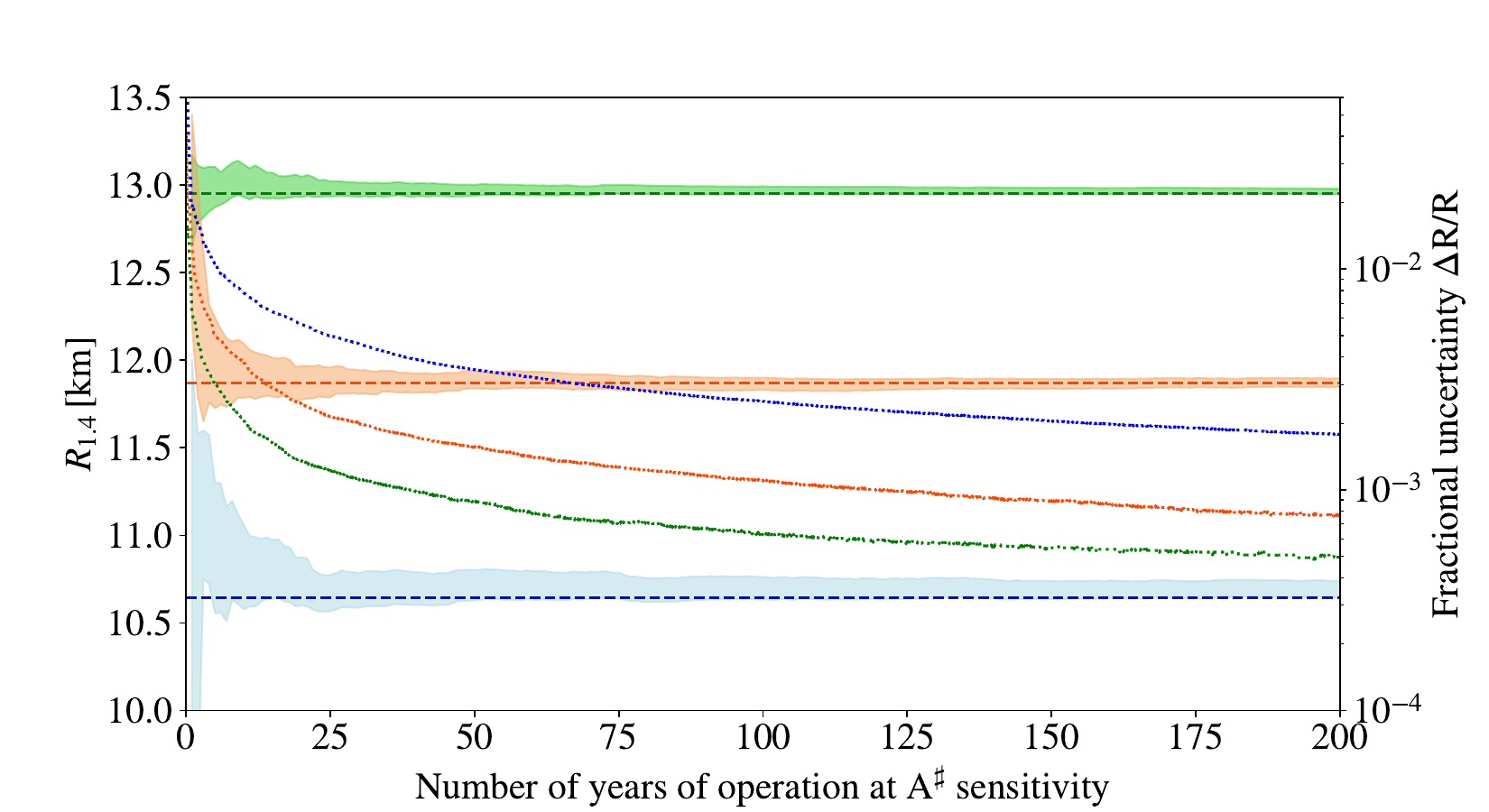}
    \\ \vspace{1cm}
    \includegraphics[height=6cm,width=0.495\linewidth]{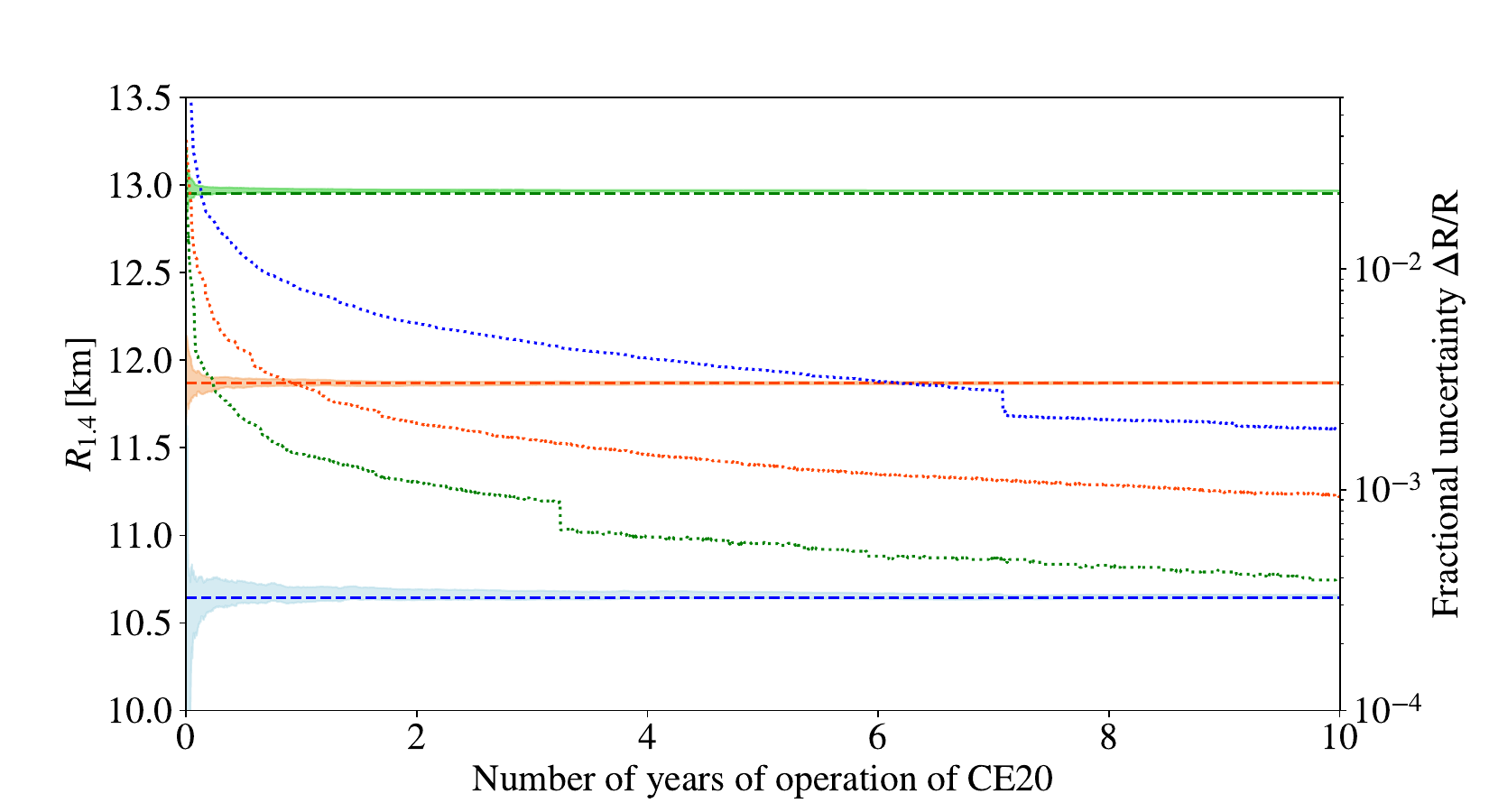}\hfill
    \includegraphics[height=6cm,width=0.495\linewidth]{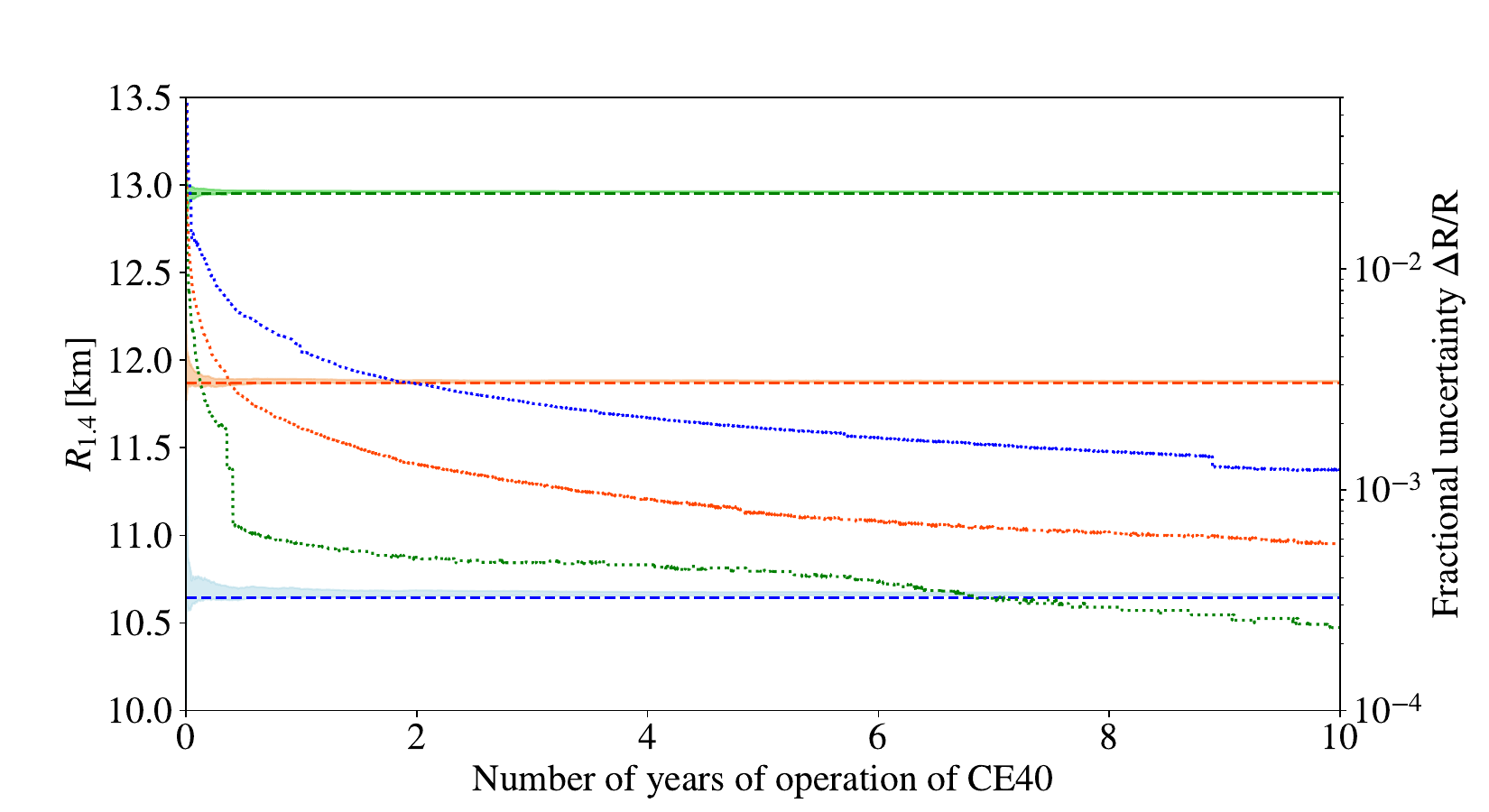}
    \\ \vspace{1cm}
    \includegraphics[height=6cm,width=0.495\linewidth]{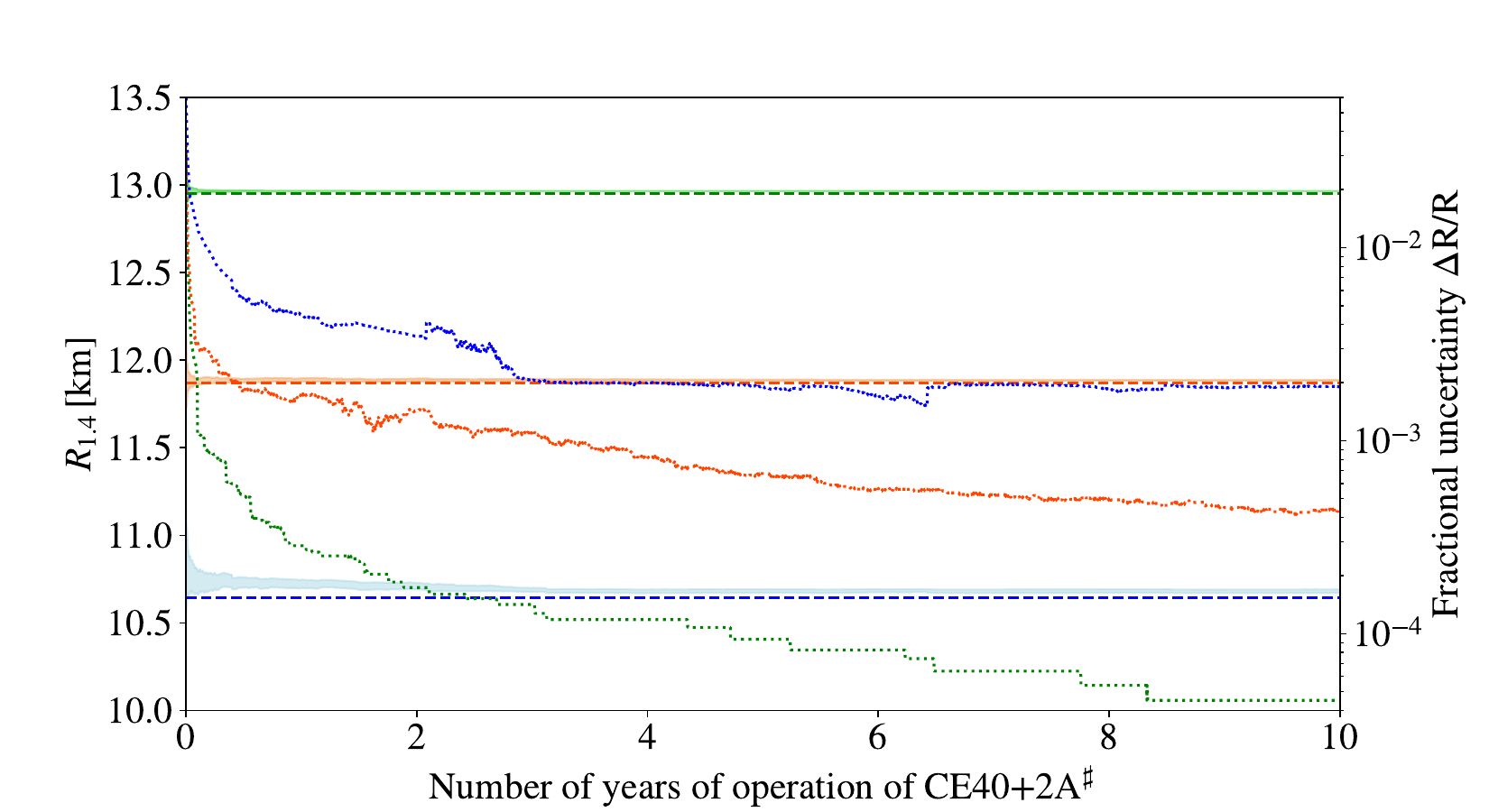}\hfill
    \includegraphics[height=6cm,width=0.495\linewidth]{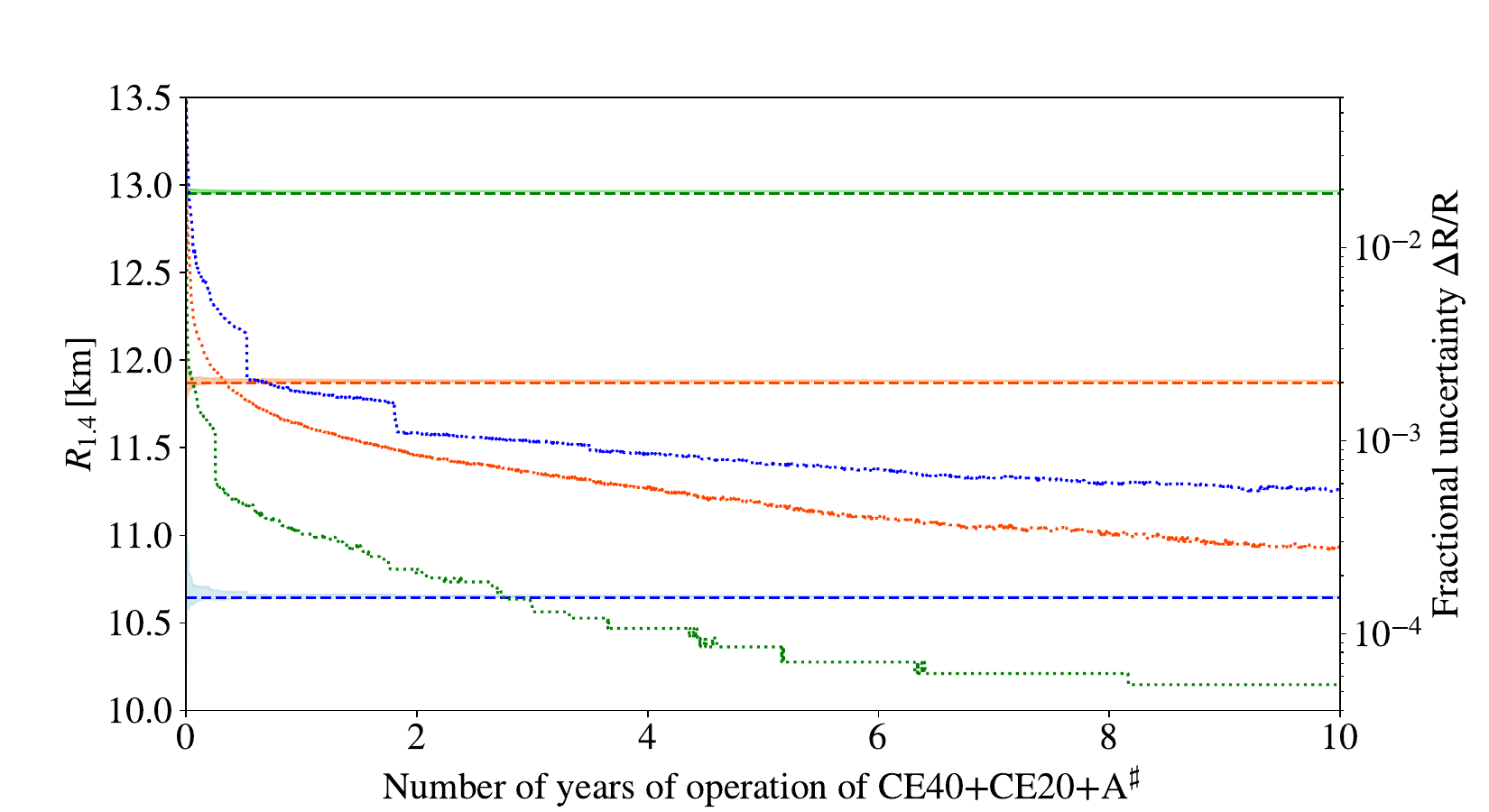}
    \caption{Combined constraints on the measurement of $R_{1.4}$ for the six different detector networks considered in this study. The blue, orange and green colors represent the constraints for the soft, medium and stiff equations of state respectively. The shaded regions in the figures represent the 90$\%$ credible intervals for the combined radius posteriors obtained using Eqn.~\ref{combinedposteriors}. The dotted lines depict the fractional error bars on radius measurement after a given number of years of observation. The dashed lines are the injected radius values. }
     \label{combined_radius_posteriors}
\end{figure*}

\begin{figure*}
    \advance\leftskip-1cm
    \includegraphics[height=6cm,width=0.52\linewidth]{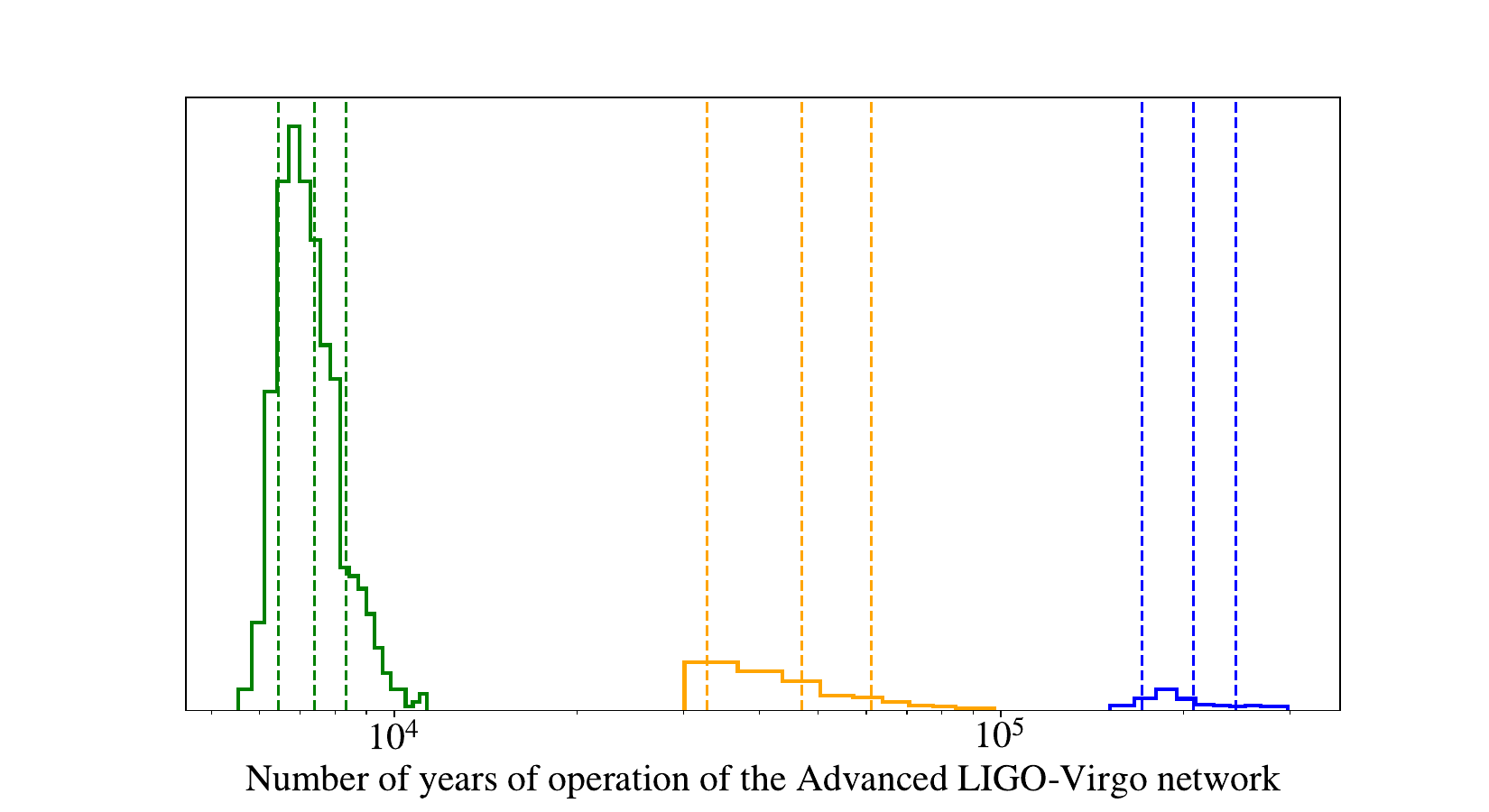}\hfill
    \includegraphics[height=6cm,width=0.52\linewidth]{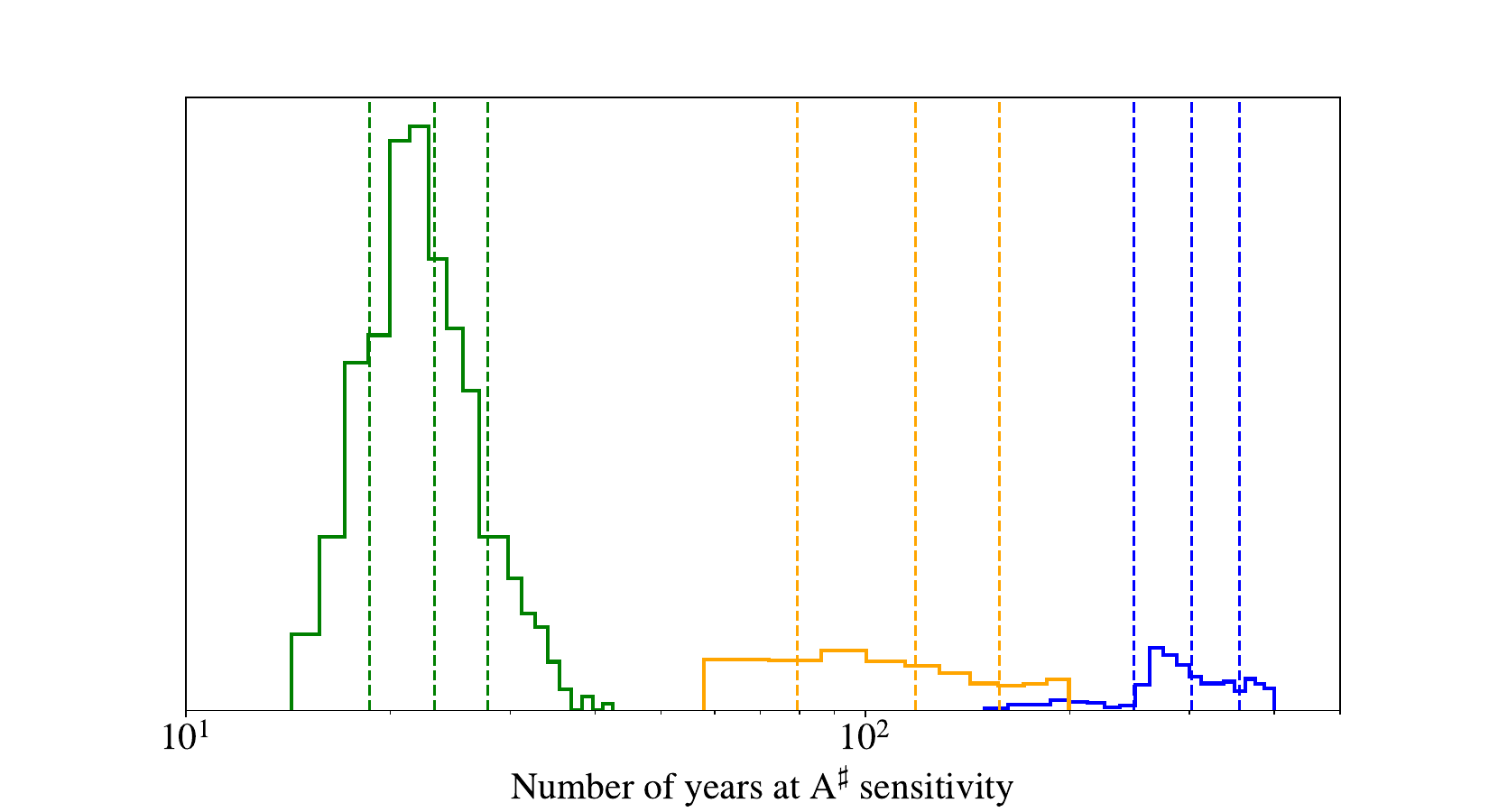}
    \\ \vspace{1cm}
    \includegraphics[height=6cm,width=0.52\linewidth]{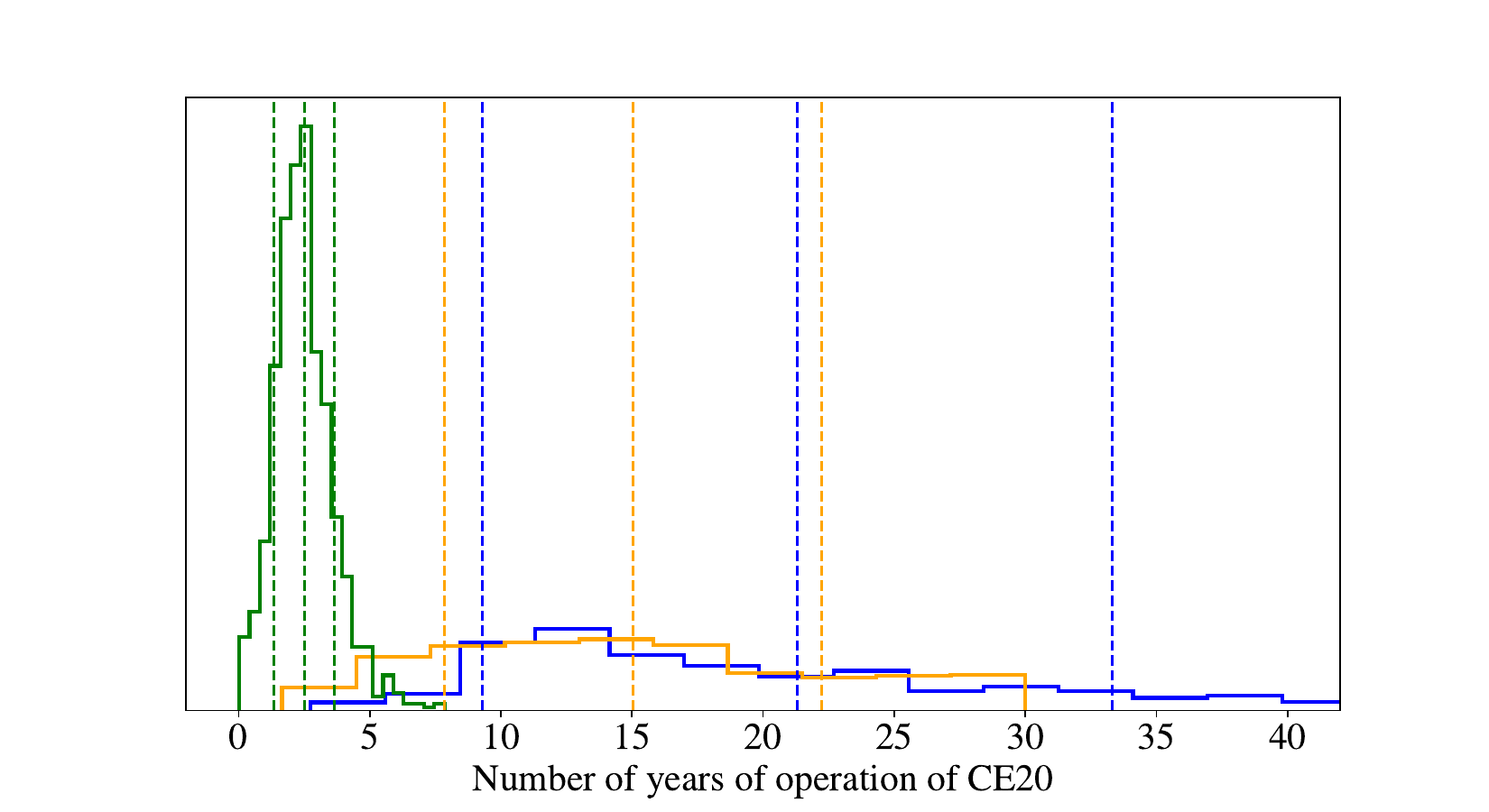}\hfill
    \includegraphics[height=6cm,width=0.52\linewidth]{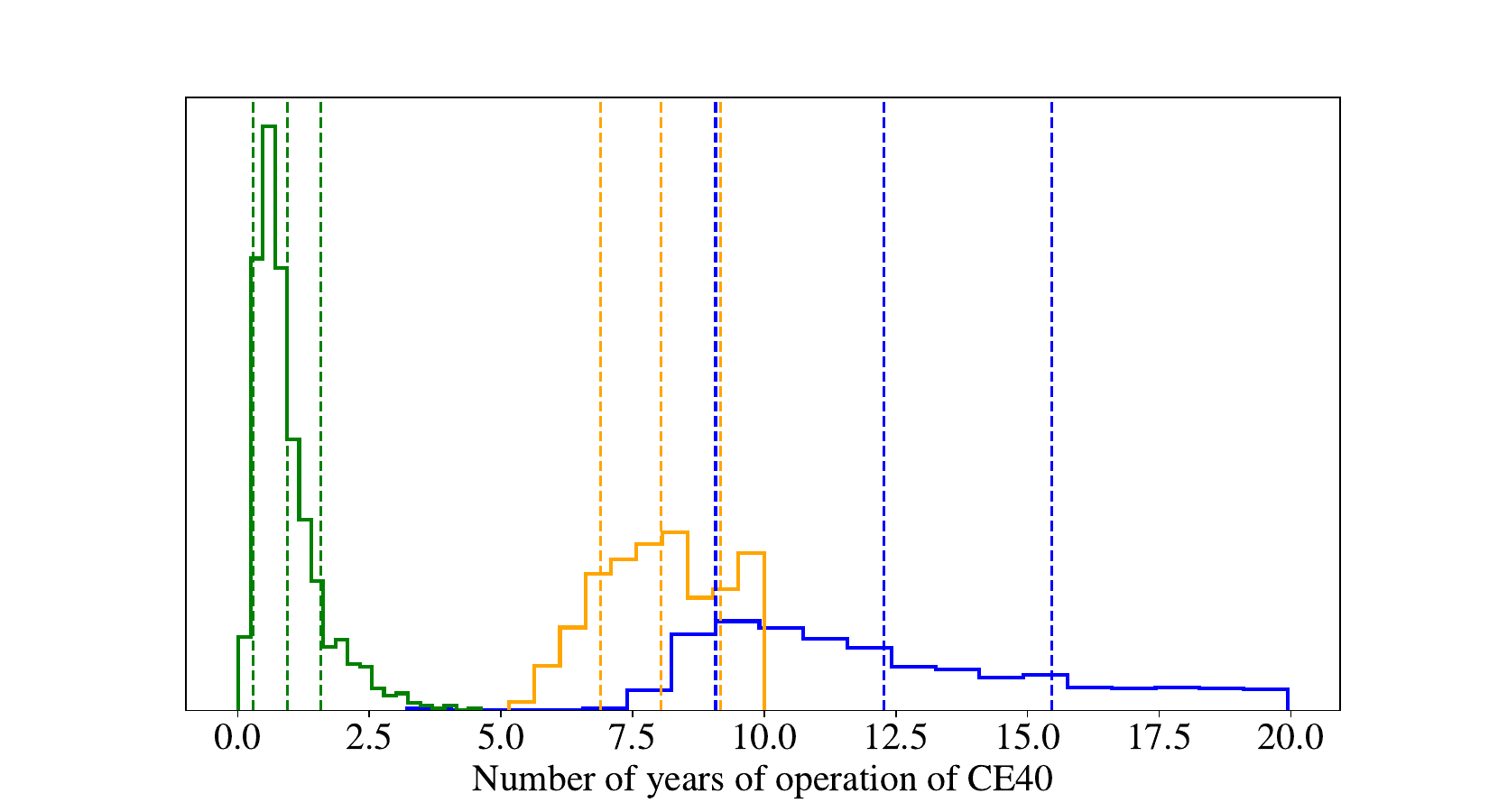}
    \\ \vspace{1cm}
    \includegraphics[height=6cm,width=0.52\linewidth]{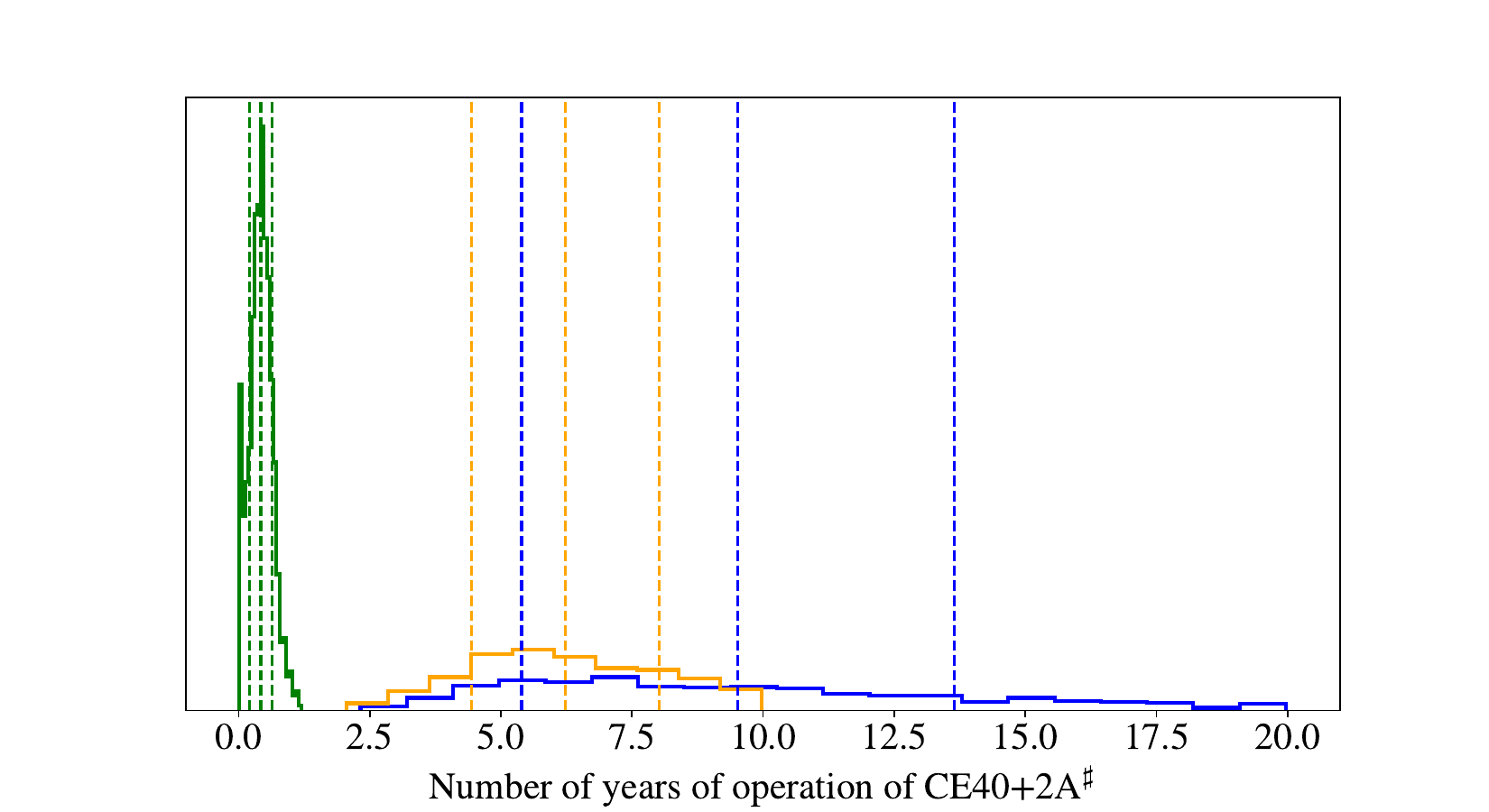}\hfill
    \includegraphics[height=6cm,width=0.52\linewidth]{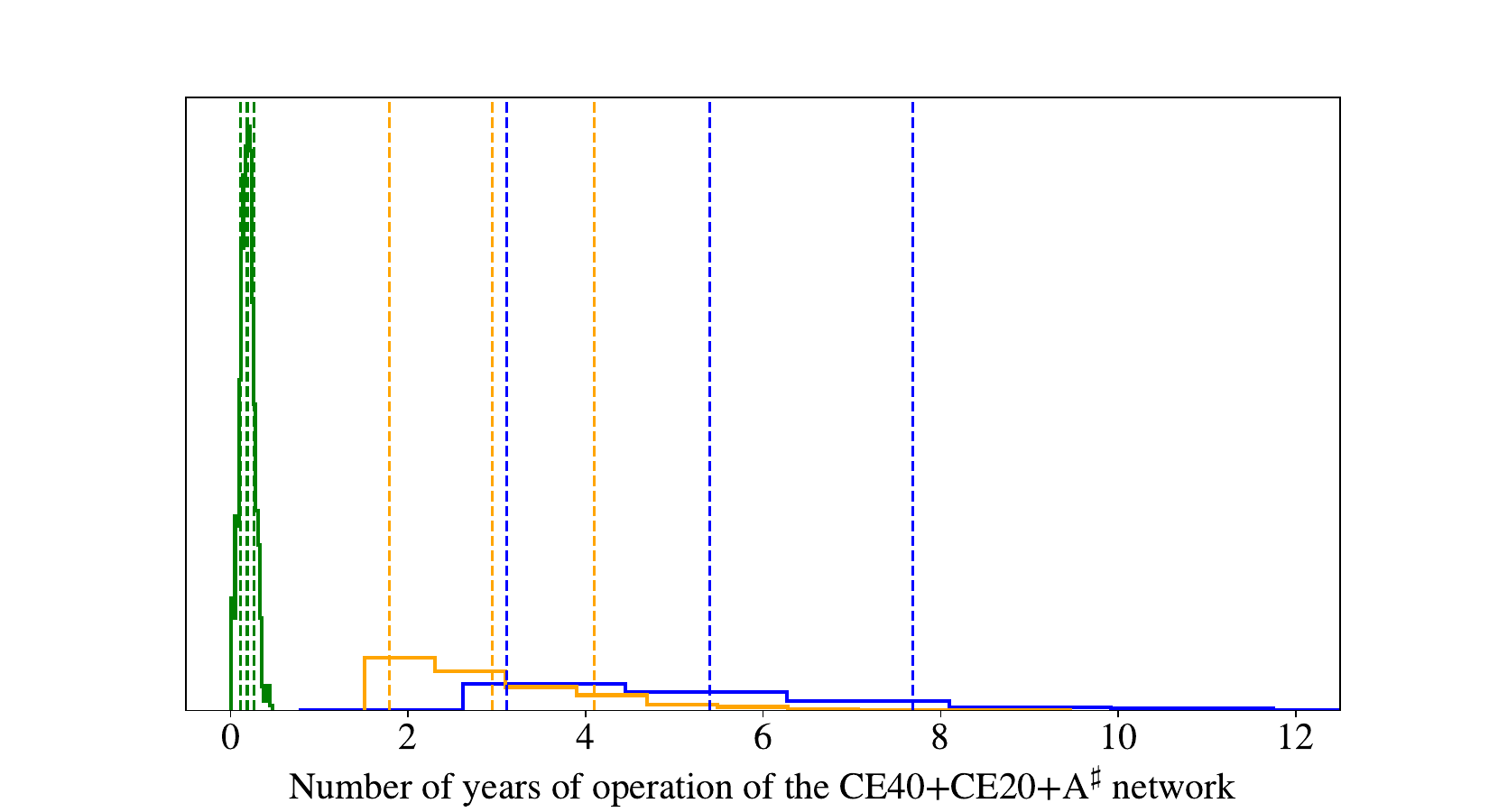}
    \caption{Distribution of the number of years of observation at different network sensitivities required to constrain the radius of a $1.4 \, \mathrm{M}_{\odot}$ neutron star down to an accuracy of 10m. The blue, orange and green histograms depict the expected distributions for the soft, medium and stiff equations of state respectively. The vertical dashed lines represent the median and $90\%$ credible intervals for each histogram.}
    \label{events_dist}
\end{figure*}

\section{Results} \label{sec:results}
We simulate a realistic observing scenario for binary neutron star mergers by creating a population of neutron stars distributed uniformly in distance (details in Section \ref{sec:population}) and re-weighting our events to a population distributed uniformly in volume. We then make random draws from this population to construct different realizations of the universe. We also randomize the sequence of detection of the events and combine the posteriors for $\rm R_{1.4}$ using Eq.~\ref{combinedposteriors} to obtain the projected accuracy in the measurement of neutron star radius for a given observation period for each detector network. In Fig.~\ref{combined_radius_posteriors}, we show the projected constraints on the radius of a $1.4 \rm M_{\odot}$ neutron star for a given random drawn population for the four detector networks described in Section~\ref{sec:methods}.  For all of the detector networks, we find that the combined radius posterior converges towards the injected value. We find that a next generation detector configuration with one 40km Cosmic Explorer, one 20km Cosmic Explorer and one A$^{\sharp}$ detector converges to the injected radius the fastest. Even one Cosmic Explorer detector converges faster to the radius than a three-detector Advanced LIGO-Virgo configuration. 

As pointed out in~\cite{Finstad23}, the loudest events have the most significant contribution towards lowering the fractional error bars on radius measurement. Thus, the exact time required to measure the radius of a neutron star to a certain accuracy will be affected by the order of occurrence of the events. To mitigate the effect of the ordering of events used in obtaining the projected constraints in Fig.~\ref{combined_radius_posteriors}, we draw different populations of the universe for each detector network 500 times and plot the distribution of the number of years of operation it would take to constrain the radius of a $1.4 \, \mathrm{M}_{\odot}$ neutron star to an accuracy of~$10 \, \rm m$. These results are shown in Fig.~\ref{events_dist} and a summary is also shown in Table~\ref{tab:table1}. From these, we can see that even for the stiffest equation of state, the current generation of ground based detectors like the Advanced LIGO-Virgo network would take several thousand years to measure the radius of a neutron star to an accuracy of 10m, and this improves only slightly with the A$^\sharp$ upgrade. For a very stiff equation of state which gives a radius of $~\sim 13 \, \rm km$ for a $1.4 \, \rm M_{\odot}$ neutron star, the A$^\sharp$ network may be able to constrain radius to an accuracy of 10m within ~20 years, but this worsens significantly if the true equation of state is moderately soft. On the other hand, a single 40km Cosmic Explorer detector will be able to constrain radius to an accuracy of 10m within 12 years for the softest equation of state considered here, and within a year for the stiffest. In order to measure the equation of state for a neutron star in a viable time-frame, building Cosmic Explorer is absolutely essential.

\begin{table*}
\begin{center}
\begin{tabular}{ccccc}
\br
Detector network &Soft EoS&Medium EoS&  Stiff EoS \\ \hline
 LIGO-Virgo&$2\rm e5 _{-4e4}^{+4e4}$&$5 \rm e4_{-1e4}^{+1e4}$&$7000_{-900}^{+900}$ \\
 $3 \, \rm A^{\sharp}$&$300_{-50}^{+50}$ &$100_{-40}^{+40}$&$20_{-4}^{+4}$\\
 CE20 &$21_{-10}^{+10}$&$15_{-7}^{+7}$&$3_{-1}^{+1}$\\
 CE40 &$12_{-3}^{+3}$&$8_{-1}^{+1}$&$1_{-0.6}^{+0.6}$\\
 CE40+$2 \, \rm A^{\sharp}$&$9_{-4}^{+4}$&$6_{-2}^{+2}$&$0.4_{-0.2}^{+0.2}$\\
 CE40+CE20+$\rm A^{\sharp}$&$5_{-2}^{+2}$ &$3_{-1}^{+1}$ &$0.2_{-0.07}^{+0.07}$\\
 \br
\end{tabular}
\caption{\label{tab:table1} A summary of the result of drawing 500 random realizations of the universe from our simulated population for each detector network and each equation of state.  For each realization of the universe, the events are shuffled to randomize order of occurrence, and then used to determine the average number of years required to measure the radius of a $1.4 \, \rm M_{\odot}$ neutron star to an accuracy of 10m.}
\end{center}
\end{table*}

\section{Conclusion} \label{sec:conclusion}
In this paper, we have performed a comparative study of the potential of different second and third generation gravitational wave detector networks to precisely constrain the nuclear equation of state.  We perform a full Bayesian inference on a simulated population of neutron stars to be observed at different network sensitivities. The equation of state models used in the analysis span the plausible range of stiffness allowed by current observational constraints for densities relevant for neutron stars, and are calibrated with chiral effective field theory at the low density regime. We combine constraints from individual events, and show that this procedure can be used to obtain a stringent constraint on the radius of a $1.4 \, \rm M_{\odot}$ neutron star.\\
Our results demonstrate the systematic improvement in the measurability of the nuclear equation of state with improvement in sensitivity of gravitational detectors, from the current generation of Advanced LIGO-Virgo detectors, to the A$^\sharp$ upgrade, and subsequently the next generation of detectors like Cosmic Explorer. The constraints obtained on the equation of state follow the expected hierarchy of measurability, with the stiffest equations of state giving the most stringent constraints, and the softer models leading to weaker constraints. Based on the current estimate for the merger rate from~\cite{LIGOScientific:2020kqk}, we find that the Advanced LIGO-Virgo detectors operating at design sensitivity would take $2\rm e5 _{-4e4}^{+4e4}$, $5 \rm e4_{-1e4}^{+1e4}$ and $7000_{-900}^{+900}$ years respectively to constrain the radius of a $1.4 \, \rm M_{\odot}$ to an accuracy of 10m, for the soft, medium and stiff equations of state respectively. With the detectors operating at A$^\sharp$ sensitivity, the same accuracy can be reached in $300_{-50}^{+50}$, $100_{-40}^{+40}$ and $20_{-4}^{+4}$ years. However, with the next generation of ground based detectors like the Cosmic Explorer, the radius of a $1.4 \, \rm M_{\odot}$ neutron star can be measured to an accuracy of 10m well within a human lifetime, with a single 20km detector doing it in $21_{-10}^{+10}$, $15_{-7}^{+7}$ and $3_{-1}^{+1}$ years, and a 40km detector requiring only $12_{-3}^{+3}$, $8_{-1}^{+1}$ and $1_{-0.6}^{+0.6}$ years respectively for the soft, medium and stiff equations of state. The multi-detector network comprised of a 40km Cosmic Explorer and 2 detectors at $\rm \Asharp$ does it in $9_{-4}^{+4}$, $6_{-2}^{+2}$ and $0.4_{-0.2}^{+0.2}$ years, whereas the network comprised of a 40km Cosmic Explorer, a 20km Cosmic Explorer and an $\rm \Asharp$ takes $5_{-2}^{+2}$, $3_{-1}^{+1}$ and $0.2_{-0.07}^{+0.07}$ years respectively to measure the radius to an accuracy of 10m for the soft, medium and stiff equations of state. Thus, the next generation of ground based detectors will be instrumental in constraining the nuclear equation of state to a level of accuracy that can not be reached with any existing facilities.

Some of the source bias in the process of obtaining combined constraints by this method are pointed out in~\cite{Finstad23}. These biases include the imperfect knowledge of the mass distribution of neutron stars and the potential absence of an electromagnetic counterpart, posing challenges in distinguishing a binary neutron star system from a neutron star black hole system. In this study, we circumvent the bias caused by the imperfect knowledge of mass distribution by choosing the mass prior used in our analysis to be the same as the one used to simulate the neutron star population. However, in practical applications, as more and more signals from binary neutron star mergers are detected, the mass distribution used in the inference should also be simultaneously updated to mitigate this bias. Another source of bias in obtaining constraints on the equation of state, particularly in terms of the neutron star radius, is that we do not measure radius directly from gravitational wave signals, but use the measurement of tidal deformability to obtain an estimate of the radius. Since the love number $k_2$ (as defined in e.g., ~\cite{Damour:2009vw,Flanagan:2007ix,Hinderer:2009ca}), relating tidal deformability to radius, is also a function of radius, even an extremely precise measurement of the tidal deformability leads to a multimodal distribution for measured radius, which can ultimately tend to bias the median value of the measured radius away from its true value. Future works can extend this work to constrain the radius not only for a $1.4 \, \rm M_{\odot}$ neutron star but for the whole mass range allowed by plausible equations of state. In conclusion, our work highlights the imperative need to build Cosmic Explorer in order to be able to accurately constrain the radius of a neutron star within a human lifetime.

Supporting data for this manuscript is available at \href{https://github.com/sugwg/bns-eos-nggw}{github.com/sugwg/bns-eos-nggw}.

\section*{Acknowledgements}
The authors acknowledge support from National Science Foundation Awards PHY-2011655 (AB and DAB), PHY-2207264 (DAB), and PHY-2309240 (AHN, KK and DAB). RS acknowledges support from the Nuclear Physics from Multi-Messenger Mergers (NP3M) Focused Research Hub which is funded by the National Science Foundation under Grant Number 21-16686, and by the Laboratory Directed Research and Development program of Los Alamos National Laboratory under project number 20220541ECR. This research was supported through computational resources provided by Syracuse University.

\printbibliography

@article{LIGOScientific:2016wof,
    author = "Abbott, Benjamin P and others",
    collaboration = "LIGO Scientific",
    title = "{Exploring the Sensitivity of Next Generation Gravitational Wave Detectors}",
    eprint = "1607.08697",
    archivePrefix = "arXiv",
    primaryClass = "astro-ph.IM",
    reportNumber = "LIGO-P1600143",
    doi = "10.1088/1361-6382/aa51f4",
    journal = "Class. Quant. Grav.",
    volume = "34",
    number = "4",
    pages = "044001",
    year = "2017"
}

@ARTICLE{2021arXiv210909882E,
       author = {{Evans}, Matthew and {Adhikari}, Rana X and {Afle}, Chaitanya and {Ballmer}, Stefan W. and {Biscoveanu}, Sylvia and {Borhanian}, Ssohrab and {Brown}, Duncan A. and {Chen}, Yanbei and {Eisenstein}, Robert and {Gruson}, Alexandra and {Gupta}, Anuradha and {Hall}, Evan D. and {Huxford}, Rachael and {Kamai}, Brittany and {Kashyap}, Rahul and {Kissel}, Jeff S. and {Kuns}, Kevin and {Landry}, Philippe and {Lenon}, Amber and {Lovelace}, Geoffrey and {McCuller}, Lee and {Ng}, Ken K.~Y. and {Nitz}, Alexander H. and {Read}, Jocelyn and {Sathyaprakash}, B.~S. and {Shoemaker}, David H. and {Slagmolen}, Bram J.~J. and {Smith}, Joshua R. and {Srivastava}, Varun and {Sun}, Ling and {Vitale}, Salvatore and {Weiss}, Rainer},
        title = "{A Horizon Study for Cosmic Explorer: Science, Observatories, and Community}",
      journal = {arXiv e-prints},
         year = 2021,
        month = sep,
          eid = {arXiv:2109.09882},
        pages = {arXiv:2109.09882},
          doi = {10.48550/arXiv.2109.09882},
archivePrefix = {arXiv},
       eprint = {2109.09882},
 primaryClass = {astro-ph.IM},
       adsurl = {https://ui.adsabs.harvard.edu/abs/2021arXiv210909882E}
}

@article{Hild:2009ns,
    author = "Hild, Stefan and Chelkowski, Simon and Freise, Andreas and Franc, Janyce and Morgado, Nazario and Flaminio, Raffaele and DeSalvo, Riccardo",
    title = "{A Xylophone Configuration for a third Generation Gravitational Wave Detector}",
    eprint = "0906.2655",
    archivePrefix = "arXiv",
    primaryClass = "gr-qc",
    doi = "10.1088/0264-9381/27/1/015003",
    journal = "Class. Quant. Grav.",
    volume = "27",
    pages = "015003",
    year = "2010"
}

@article{Sathyaprakash:2012jk,
    author = "Sathyaprakash, B. and others",
    editor = "Hannam, Mark and Sutton, Patrick and Hild, Stefan and van den Broeck, Chris",
    title = "{Scientific Objectives of Einstein Telescope}",
    eprint = "1206.0331",
    archivePrefix = "arXiv",
    primaryClass = "gr-qc",
    doi = "10.1088/0264-9381/29/12/124013",
    journal = "Class. Quant. Grav.",
    volume = "29",
    pages = "124013",
    year = "2012",
    note = "[Erratum: Class.Quant.Grav. 30, 079501 (2013)]"
}

@article{Damour:2009vw,
    author = "Damour, Thibault and Nagar, Alessandro",
    title = "{Relativistic tidal properties of neutron stars}",
    eprint = "0906.0096",
    archivePrefix = "arXiv",
    primaryClass = "gr-qc",
    doi = "10.1103/PhysRevD.80.084035",
    journal = "Phys. Rev. D",
    volume = "80",
    pages = "084035",
    year = "2009"
}

@article{Flanagan:2007ix,
    author = "Flanagan, Eanna E. and Hinderer, Tanja",
    title = "{Constraining neutron star tidal Love numbers with gravitational wave detectors}",
    eprint = "0709.1915",
    archivePrefix = "arXiv",
    primaryClass = "astro-ph",
    doi = "10.1103/PhysRevD.77.021502",
    journal = "Phys. Rev. D",
    volume = "77",
    pages = "021502",
    year = "2008"
}

@article{Hinderer:2009ca,
    author = "Hinderer, Tanja and Lackey, Benjamin D. and Lang, Ryan N. and Read, Jocelyn S.",
    title = "{Tidal deformability of neutron stars with realistic equations of state and their gravitational wave signatures in binary inspiral}",
    eprint = "0911.3535",
    archivePrefix = "arXiv",
    primaryClass = "astro-ph.HE",
    doi = "10.1103/PhysRevD.81.123016",
    journal = "Phys. Rev. D",
    volume = "81",
    pages = "123016",
    year = "2010"
}

@article{Lackey:2014fwa,
    author = "Lackey, Benjamin D. and Wade, Leslie",
    title = "{Reconstructing the neutron-star equation of state with gravitational-wave detectors from a realistic population of inspiralling binary neutron stars}",
    eprint = "1410.8866",
    archivePrefix = "arXiv",
    primaryClass = "gr-qc",
    doi = "10.1103/PhysRevD.91.043002",
    journal = "Phys. Rev. D",
    volume = "91",
    number = "4",
    pages = "043002",
    year = "2015"
}

@article{Agathos:2015uaa,
    author = "Agathos, Michalis and Meidam, Jeroen and Del Pozzo, Walter and Li, Tjonnie G. F. and Tompitak, Marco and Veitch, John and Vitale, Salvatore and Van Den Broeck, Chris",
    title = "{Constraining the neutron star equation of state with gravitational wave signals from coalescing binary neutron stars}",
    eprint = "1503.05405",
    archivePrefix = "arXiv",
    primaryClass = "gr-qc",
    doi = "10.1103/PhysRevD.92.023012",
    journal = "Phys. Rev. D",
    volume = "92",
    number = "2",
    pages = "023012",
    year = "2015"
}

@article{HernandezVivanco:2019vvk,
    author = "Hernandez Vivanco, Francisco and Smith, Rory and Thrane, Eric and Lasky, Paul D. and Talbot, Colm and Raymond, Vivien",
    title = "{Measuring the neutron star equation of state with gravitational waves: The first forty binary neutron star merger observations}",
    eprint = "1909.02698",
    archivePrefix = "arXiv",
    primaryClass = "gr-qc",
    doi = "10.1103/PhysRevD.100.103009",
    journal = "Phys. Rev. D",
    volume = "100",
    number = "10",
    pages = "103009",
    year = "2019"
}

@article{Pacilio:2021jmq,
    author = "Pacilio, Costantino and Maselli, Andrea and Fasano, Margherita and Pani, Paolo",
    title = "{Ranking Love Numbers for the Neutron Star Equation of State: The Need for Third-Generation Detectors}",
    eprint = "2104.10035",
    archivePrefix = "arXiv",
    primaryClass = "gr-qc",
    doi = "10.1103/PhysRevLett.128.101101",
    journal = "Phys. Rev. Lett.",
    volume = "128",
    number = "10",
    pages = "101101",
    year = "2022"
}

@article{Chatziioannou:2021tdi,
    author = "Chatziioannou, Katerina",
    title = "{Uncertainty limits on neutron star radius measurements with gravitational waves}",
    eprint = "2108.12368",
    archivePrefix = "arXiv",
    primaryClass = "gr-qc",
    doi = "10.1103/PhysRevD.105.084021",
    journal = "Phys. Rev. D",
    volume = "105",
    number = "8",
    pages = "084021",
    year = "2022"
}

@article{Capano:2019eae,
    author = "Capano, Collin D. and Tews, Ingo and Brown, Stephanie M. and Margalit, Ben and De, Soumi and Kumar, Sumit and Brown, Duncan A. and Krishnan, Badri and Reddy, Sanjay",
    title = "{Stringent constraints on neutron-star radii from multimessenger observations and nuclear theory}",
    eprint = "1908.10352",
    archivePrefix = "arXiv",
    primaryClass = "astro-ph.HE",
    reportNumber = "INT-PUB-19-037, LA-UR-19-28442",
    doi = "10.1038/s41550-020-1014-6",
    journal = "Nature Astron.",
    volume = "4",
    number = "6",
    pages = "625--632",
    year = "2020"
}

@article{Husa:2015iqa,
    author = {Husa, Sascha and Khan, Sebastian and Hannam, Mark and P\"urrer, Michael and Ohme, Frank and Jim\'enez Forteza, Xisco and Boh\'e, Alejandro},
    title = "{Frequency-domain gravitational waves from nonprecessing black-hole binaries. I. New numerical waveforms and anatomy of the signal}",
    eprint = "1508.07250",
    archivePrefix = "arXiv",
    primaryClass = "gr-qc",
    doi = "10.1103/PhysRevD.93.044006",
    journal = "Phys. Rev. D",
    volume = "93",
    number = "4",
    pages = "044006",
    year = "2016"
}

@article{Khan:2015jqa,
    author = {Khan, Sebastian and Husa, Sascha and Hannam, Mark and Ohme, Frank and P\"urrer, Michael and Jim\'enez Forteza, Xisco and Boh\'e, Alejandro},
    title = "{Frequency-domain gravitational waves from nonprecessing black-hole binaries. II. A phenomenological model for the advanced detector era}",
    eprint = "1508.07253",
    archivePrefix = "arXiv",
    primaryClass = "gr-qc",
    doi = "10.1103/PhysRevD.93.044007",
    journal = "Phys. Rev. D",
    volume = "93",
    number = "4",
    pages = "044007",
    year = "2016"
}

@article{Barkett:2015wia,
    author = "Barkett, Kevin and others",
    title = "{Gravitational waveforms for neutron star binaries from binary black hole simulations}",
    eprint = "1509.05782",
    archivePrefix = "arXiv",
    primaryClass = "gr-qc",
    doi = "10.1103/PhysRevD.93.044064",
    journal = "Phys. Rev. D",
    volume = "93",
    number = "4",
    pages = "044064",
    year = "2016"
}

@article{Dietrich:2017aum,
    author = "Dietrich, Tim and Bernuzzi, Sebastiano and Tichy, Wolfgang",
    title = "{Closed-form tidal approximants for binary neutron star gravitational waveforms constructed from high-resolution numerical relativity simulations}",
    eprint = "1706.02969",
    archivePrefix = "arXiv",
    primaryClass = "gr-qc",
    doi = "10.1103/PhysRevD.96.121501",
    journal = "Phys. Rev. D",
    volume = "96",
    number = "12",
    pages = "121501",
    year = "2017"
}

@article{Biwer:2018osg,
    author = "Biwer, C. M. and Capano, Collin D. and De, Soumi and Cabero, Miriam and Brown, Duncan A. and Nitz, Alexander H. and Raymond, V.",
    title = "{PyCBC Inference: A Python-based parameter estimation toolkit for compact binary coalescence signals}",
    eprint = "1807.10312",
    archivePrefix = "arXiv",
    primaryClass = "astro-ph.IM",
    doi = "10.1088/1538-3873/aaef0b",
    journal = "Publ. Astron. Soc. Pac.",
    volume = "131",
    number = "996",
    pages = "024503",
    year = "2019"
}

@article{Foreman-Mackey:2012any,
    author = "Foreman-Mackey, Daniel and Hogg, David W. and Lang, Dustin and Goodman, Jonathan",
    title = "{emcee: The MCMC Hammer}",
    eprint = "1202.3665",
    archivePrefix = "arXiv",
    primaryClass = "astro-ph.IM",
    doi = "10.1086/670067",
    journal = "Publ. Astron. Soc. Pac.",
    volume = "125",
    pages = "306--312",
    year = "2013"
}

@BOOK{1970esn..book.....W,
       author = {{Wainstein}, L.~A. and {Zubakov}, V.~D.},
        title = "{Extraction of Signals from Noise}",
         year = 1970,
       adsurl = {https://ui.adsabs.harvard.edu/abs/1970esn..book.....W}
}

@article{Zackay:2018qdy,
    author = "Zackay, Barak and Dai, Liang and Venumadhav, Tejaswi",
    title = "{Relative Binning and Fast Likelihood Evaluation for Gravitational Wave Parameter Estimation}",
    eprint = "1806.08792",
    archivePrefix = "arXiv",
    primaryClass = "astro-ph.IM",
    month = "6",
    year = "2018"
}

@article{Finstad:2020sok,
    author = "Finstad, Daniel and Brown, Duncan A.",
    title = "{Fast Parameter Estimation of Binary Mergers for Multimessenger Follow-up}",
    eprint = "2009.13759",
    archivePrefix = "arXiv",
    primaryClass = "astro-ph.IM",
    doi = "10.3847/2041-8213/abca9e",
    journal = "Astrophys. J. Lett.",
    volume = "905",
    number = "1",
    pages = "L9",
    year = "2020"
}

@article{10.1214/aoms/1177728190,
author = {Murray Rosenblatt},
title = {{Remarks on Some Nonparametric Estimates of a Density Function}},
volume = {27},
journal = {The Annals of Mathematical Statistics},
number = {3},
publisher = {Institute of Mathematical Statistics},
pages = {832 -- 837},
year = {1956},
doi = {10.1214/aoms/1177728190},
URL = {https://doi.org/10.1214/aoms/1177728190}
}

@article{LIGOScientific:2020kqk,
    author = "Abbott, R. and others",
    collaboration = "LIGO Scientific, Virgo",
    title = "{Population Properties of Compact Objects from the Second LIGO-Virgo Gravitational-Wave Transient Catalog}",
    eprint = "2010.14533",
    archivePrefix = "arXiv",
    primaryClass = "astro-ph.HE",
    reportNumber = "LIGO-P2000077",
    doi = "10.3847/2041-8213/abe949",
    journal = "Astrophys. J. Lett.",
    volume = "913",
    number = "1",
    pages = "L7",
    year = "2021"
}

@article{KAGRA:2013rdx,
    author = "Abbott, B. P. and others",
    collaboration = "KAGRA, LIGO Scientific, Virgo, VIRGO",
    title = "{Prospects for observing and localizing gravitational-wave transients with Advanced LIGO, Advanced Virgo and KAGRA}",
    eprint = "1304.0670",
    archivePrefix = "arXiv",
    primaryClass = "gr-qc",
    reportNumber = "LIGO-P1200087, VIR-0288A-12, JGW-P1808427",
    doi = "10.1007/s41114-020-00026-9",
    journal = "Living Rev. Rel.",
    volume = "21",
    number = "1",
    pages = "3",
    year = "2018"
}

@article{LIGOScientific:2016emj,
    author = "Abbott, B. P. and others",
    collaboration = "LIGO Scientific, Virgo",
    title = "{GW150914: The Advanced LIGO Detectors in the Era of First Discoveries}",
    eprint = "1602.03838",
    archivePrefix = "arXiv",
    primaryClass = "gr-qc",
    reportNumber = "LIGO-P1500237",
    doi = "10.1103/PhysRevLett.116.131103",
    journal = "Phys. Rev. Lett.",
    volume = "116",
    number = "13",
    pages = "131103",
    year = "2016"
}

@ARTICLE{2020PhRvD.102f2003B,
       author = {{Buikema}, A. and {Cahillane}, C. and {Mansell}, G.~L. and {Blair}, C.~D. and {Abbott}, R. and {Adams}, C. and {Adhikari}, R.~X. and {Ananyeva}, A. and {Appert}, S. and {Arai}, K. and {Areeda}, J.~S. and {Asali}, Y. and {Aston}, S.~M. and {Austin}, C. and {Baer}, A.~M. and {Ball}, M. and {Ballmer}, S.~W. and {Banagiri}, S. and {Barker}, D. and {Barsotti}, L. and {Bartlett}, J. and {Berger}, B.~K. and {Betzwieser}, J. and {Bhattacharjee}, D. and {Billingsley}, G. and {Biscans}, S. and {Blair}, R.~M. and {Bode}, N. and {Booker}, P. and {Bork}, R. and {Bramley}, A. and {Brooks}, A.~F. and {Brown}, D.~D. and {Cannon}, K.~C. and {Chen}, X. and {Ciobanu}, A.~A. and {Clara}, F. and {Cooper}, S.~J. and {Corley}, K.~R. and {Countryman}, S.~T. and {Covas}, P.~B. and {Coyne}, D.~C. and {Datrier}, L.~E.~H. and {Davis}, D. and {Di Fronzo}, C. and {Dooley}, K.~L. and {Driggers}, J.~C. and {Dupej}, P. and {Dwyer}, S.~E. and {Effler}, A. and {Etzel}, T. and {Evans}, M. and {Evans}, T.~M. and {Feicht}, J. and {Fernandez-Galiana}, A. and {Fritschel}, P. and {Frolov}, V.~V. and {Fulda}, P. and {Fyffe}, M. and {Giaime}, J.~A. and {Giardina}, K.~D. and {Godwin}, P. and {Goetz}, E. and {Gras}, S. and {Gray}, C. and {Gray}, R. and {Green}, A.~C. and {Gustafson}, E.~K. and {Gustafson}, R. and {Hanks}, J. and {Hanson}, J. and {Hardwick}, T. and {Hasskew}, R.~K. and {Heintze}, M.~C. and {Helmling-Cornell}, A.~F. and {Holland}, N.~A. and {Jones}, J.~D. and {Kandhasamy}, S. and {Karki}, S. and {Kasprzack}, M. and {Kawabe}, K. and {Kijbunchoo}, N. and {King}, P.~J. and {Kissel}, J.~S. and {Kumar}, Rahul and {Landry}, M. and {Lane}, B.~B. and {Lantz}, B. and {Laxen}, M. and {Lecoeuche}, Y.~K. and {Leviton}, J. and {Liu}, J. and {Lormand}, M. and {Lundgren}, A.~P. and {Macas}, R. and {MacInnis}, M. and {Macleod}, D.~M. and {M{\'a}rka}, S. and {M{\'a}rka}, Z. and {Martynov}, D.~V. and {Mason}, K. and {Massinger}, T.~J. and {Matichard}, F. and {Mavalvala}, N. and {McCarthy}, R. and {McClelland}, D.~E. and {McCormick}, S. and {McCuller}, L. and {McIver}, J. and {McRae}, T. and {Mendell}, G. and {Merfeld}, K. and {Merilh}, E.~L. and {Meylahn}, F. and {Mistry}, T. and {Mittleman}, R. and {Moreno}, G. and {Mow-Lowry}, C.~M. and {Mozzon}, S. and {Mullavey}, A. and {Nelson}, T.~J.~N. and {Nguyen}, P. and {Nuttall}, L.~K. and {Oberling}, J. and {Oram}, Richard J. and {O'Reilly}, B. and {Osthelder}, C. and {Ottaway}, D.~J. and {Overmier}, H. and {Palamos}, J.~R. and {Parker}, W. and {Payne}, E. and {Pele}, A. and {Penhorwood}, R. and {Perez}, C.~J. and {Pirello}, M. and {Radkins}, H. and {Ramirez}, K.~E. and {Richardson}, J.~W. and {Riles}, K. and {Robertson}, N.~A. and {Rollins}, J.~G. and {Romel}, C.~L. and {Romie}, J.~H. and {Ross}, M.~P. and {Ryan}, K. and {Sadecki}, T. and {Sanchez}, E.~J. and {Sanchez}, L.~E. and {Saravanan}, T.~R. and {Savage}, R.~L. and {Schaetzl}, D. and {Schnabel}, R. and {Schofield}, R.~M.~S. and {Schwartz}, E. and {Sellers}, D. and {Shaffer}, T. and {Sigg}, D. and {Slagmolen}, B.~J.~J. and {Smith}, J.~R. and {Soni}, S. and {Sorazu}, B. and {Spencer}, A.~P. and {Strain}, K.~A. and {Sun}, L. and {Szczepa{\'n}czyk}, M.~J. and {Thomas}, M. and {Thomas}, P. and {Thorne}, K.~A. and {Toland}, K. and {Torrie}, C.~I. and {Traylor}, G. and {Tse}, M. and {Urban}, A.~L. and {Vajente}, G. and {Valdes}, G. and {Vander-Hyde}, D.~C. and {Veitch}, P.~J. and {Venkateswara}, K. and {Venugopalan}, G. and {Viets}, A.~D. and {Vo}, T. and {Vorvick}, C. and {Wade}, M. and {Ward}, R.~L. and {Warner}, J. and {Weaver}, B. and {Weiss}, R. and {Whittle}, C. and {Willke}, B. and {Wipf}, C.~C. and {Xiao}, L. and {Yamamoto}, H. and {Yu}, Hang and {Yu}, Haocun and {Zhang}, L. and {Zucker}, M.~E. and {Zweizig}, J.},
        title = "{Sensitivity and performance of the Advanced LIGO detectors in the third observing run}",
      journal = {\prd},
         year = 2020,
        month = sep,
       volume = {102},
       number = {6},
          eid = {062003},
        pages = {062003},
          doi = {10.1103/PhysRevD.102.062003},
archivePrefix = {arXiv},
       eprint = {2008.01301},
 primaryClass = {astro-ph.IM},
       adsurl = {https://ui.adsabs.harvard.edu/abs/2020PhRvD.102f2003B}
}

@article{VIRGO:2014yos,
    author = "Acernese, F. and others",
    collaboration = "VIRGO",
    title = "{Advanced Virgo: a second-generation interferometric gravitational wave detector}",
    eprint = "1408.3978",
    archivePrefix = "arXiv",
    primaryClass = "gr-qc",
    doi = "10.1088/0264-9381/32/2/024001",
    journal = "Class. Quant. Grav.",
    volume = "32",
    number = "2",
    pages = "024001",
    year = "2015"
}

@techreport{T2200287,
    author = {Fritschel, Peter and Kuns, Kevin and Driggers, Jenne and Effler, Anamaria and Lantz, Brian and Ottaway, David and Ballmer, Stefan and Dooley, Kate and Adhikari, Rana and Evans, Matthew and Farr, Ben and Gonzalez, Gabriela and Schmidt, Patricia and Raja, Sendhil},
    title={Report from the LSC Post-O5 Study Group},
    institution={LIGO},
    number={T2200287},
    url={https://dcc.ligo.org/LIGO-T2200287/public},
    year={2022}
}

@article{Cornish:2021lje,
    author = "Cornish, Neil J.",
    title = "{Heterodyned likelihood for rapid gravitational wave parameter inference}",
    eprint = "2109.02728",
    archivePrefix = "arXiv",
    primaryClass = "gr-qc",
    doi = "10.1103/PhysRevD.104.104054",
    journal = "Phys. Rev. D",
    volume = "104",
    number = "10",
    pages = "104054",
    year = "2021"
}

@article{Dietrich:2020eud,
    author = "Dietrich, Tim and Hinderer, Tanja and Samajdar, Anuradha",
    title = "{Interpreting Binary Neutron Star Mergers: Describing the Binary Neutron Star Dynamics, Modelling Gravitational Waveforms, and Analyzing Detections}",
    eprint = "2004.02527",
    archivePrefix = "arXiv",
    primaryClass = "gr-qc",
    doi = "10.1007/s10714-020-02751-6",
    journal = "Gen. Rel. Grav.",
    volume = "53",
    number = "3",
    pages = "27",
    year = "2021"
}

@article{Dudi:2018jzn,
    author = {Dudi, Reetika and Pannarale, Francesco and Dietrich, Tim and Hannam, Mark and Bernuzzi, Sebastiano and Ohme, Frank and Br\"ugmann, Bernd},
    title = "{Relevance of tidal effects and post-merger dynamics for binary neutron star parameter estimation}",
    eprint = "1808.09749",
    archivePrefix = "arXiv",
    primaryClass = "gr-qc",
    doi = "10.1103/PhysRevD.98.084061",
    journal = "Phys. Rev. D",
    volume = "98",
    number = "8",
    pages = "084061",
    year = "2018"
}

@article{Samajdar:2018dcx,
    author = "Samajdar, Anuradha and Dietrich, Tim",
    title = "{Waveform systematics for binary neutron star gravitational wave signals: effects of the point-particle baseline and tidal descriptions}",
    eprint = "1810.03936",
    archivePrefix = "arXiv",
    primaryClass = "gr-qc",
    doi = "10.1103/PhysRevD.98.124030",
    journal = "Phys. Rev. D",
    volume = "98",
    number = "12",
    pages = "124030",
    year = "2018"
}

@article{Finstad23,
       author = {{Finstad}, Daniel and {White}, Laurel V. and {Brown}, Duncan A.},
        title = "{Prospects for a Precise Equation of State Measurement from Advanced LIGO and Cosmic Explorer}",
      journal = {\apj},
         year = 2023,
        month = sep,
       volume = {955},
       number = {1},
          eid = {45},
        pages = {45},
          doi = {10.3847/1538-4357/acf12f},
archivePrefix = {arXiv},
       eprint = {2211.01396},
 primaryClass = {astro-ph.HE},
       adsurl = {https://ui.adsabs.harvard.edu/abs/2023ApJ...955...45F}
}

\end{document}